\documentclass[twocolumn,twoside,slac]{revtex4}
\usepackage{graphicx}
\usepackage{fancyhdr}
\usepackage{xspace}
\newcommand{\CondorG}[0]{\mbox{Condor-G}\xspace}

\begin{document}

%

\title{Management of Grid Jobs and Information within SAMGrid}
\author{A.~Baranovski, G.~Garzoglio, A.~Kreymer, L.~Lueking, \\
S.~Stonjek, I.~Terekhov, F.~Wuerthwein
}
\affiliation{FNAL, Batavia, IL 60510, USA}
\author{A.~Roy,  T.~Tannenbaum}
\affiliation{University of Wisconsin, Madison, WI 53706 , USA}
\author{P.~Mhashilkar, V.~Murthi}
\affiliation{UTA, Arlington, TX 76019, USA}
\author{R.~Walker}
\affiliation{Imperial College, London, UK}
\author{F.~Ratnikov}
\affiliation{Rutgers University, Piscataway NJ 08854, USA}
\author{T.~Rockwell}
\affiliation{Michigan State University, East Lansing, MI  48824, USA}


\begin{abstract}
We describe some of the key aspects of the SAMGrid system, used
by the D0 and CDF experiments at Fermilab. Having sustained
success of the data handling part of SAMGrid, we have developed
new services for job and information services. Our job management is
rooted in \CondorG and uses enhancements that are general 
applicability for HEP grids. Our information system is based
on a uniform framework for configuration management based on
XML data representation and processing.
\end{abstract}

\maketitle

\thispagestyle{fancy}

\section{Introduction}

Grid \cite{grid_book} 
has emerged as a modern trend in computing, aiming to support the sharing
and coordinated use of diverse resources by Virtual Organizations (VO's)
in order to solve their common problems\cite{grid_anatomy}. 
It was originally driven by
by scientific, and especially High-Energy Physics (HEP) communities.
HEP experiments are a classic example of
large, globally distributed VO's whose participants are scattered
over many institutions and collaborate on 
studies of experimental data, primarily on data processing and analysis.

Our background is specifically
in the development of large-scale, globally distributed
systems for HEP experiments. We apply grid technologies to our systems 
and develop higher-level, community-specific grid services
(generally defined in \cite{grid_services}), currently
for the two collider experiments at Fermilab, D0 and CDF. These two
experiments are actually the largest currently running HEP experiments,
each having over half a thousand users and planning to analyze repeatedly 
peta-byte scale data. 

The success of the distributed computing for the experiments depends
on many factors. In the HEP computing, which
remains a principal application domain for the Grid as a whole, jobs
are {\it data-intensive} and therefore {\it data handling} is one of the most
important factors. For HEP experiments such as D0 and CDF, data handling
is the center of the meta-computing grid system \cite{acat2002_it}.
The SAM data handling system \cite{samwww} was originally developed
for the D0 collaboration and is currently also used by CDF. The system
is described in detail elsewhere (see, for example, 
\cite{sam_chep2001,chep2001_grid} and references therein). 
Here, we only note
some of the advanced features of the system -- the ability to coordinate
multiple concurrent accesses to Storage Systems 
\cite{hpdc2001}
and global data routing and replication \cite{acat2000}.

Given the ability to distribute data on demand globally, we face the
similar challenges of distributing the processing of the data. Generally,
for this purpose we need global job scheduling and information
management, which is a term we prefer over ``monitoring'' as
we strive to include configuration management, resource description, and
logging.

In recent years, we have been
working on the SAMGrid project \cite{acat2002_gg}, which 
addresses the grid needs of the experiments; our current focus is in the
Jobs and Information Management (JIM), which is to complement
the SAM grid data handling system with services for job submission,
brokering and execution as well as distributed monitoring. 
Together, SAM and JIM form SAMGrid, a ``VO-specific'' grid system. 

In this paper, we present some key ideas from our system's design.
For job management per se, we collaborate with the Condor team
to enhance the \CondorG middleware so as to enable scheduling of
data-intensive jobs with flexible resource description. For
information, we focus on describing the sites' resources in the tree-like
structures of XML, with subsequent projections onto the Condor 
Classified Advertisements (ClassAd)
framework, monitoring with Globus MDS and other tools.

The rest of the paper is organized as follows. We discuss the 
relevant job scheduling design issues and Condor enhancements in 
Section \ref{sec:jobs}. In Section \ref{sec:info}, we describe configuration
management and monitoring. In Section \ref{sec:status}, we present the
status of the project and interfaces with the experiments' computing
environment and fabric; we conclude in Section \ref{sec:summary}.

\section {Job Scheduling and Brokering \label{sec:jobs}}
A key area in Grid computing is job management,
which typically includes planning of job's dependencies, selection of the
execution cluster(s) for the job, scheduling of the job at the cluster(s)
and ensuring reliable submission and execution. We base our solution
on the \CondorG framework \cite{condorg}, a powerful Grid middleware
commonly used for distributed computing. Thanks to the Particle Physics
Data Grid (PPDG) collaboration \cite{ppdg}, we have been able to work
with the Condor team to enhance the \CondorG framework and then
implement higher-level functions on top of it. In this Section,
we first summarize the general \CondorG enhancements and then
proceed to actually describing how we schedule data-intensive
jobs for D0 and CDF.

\subsection{The Grid Enhancements Condor \label{sec:condor}}

We have designed three principal enhancements for \CondorG. These
have all been successfully implemented by the Condor team:

\begin{itemize}
\item Original \CondorG required users to either specify which grid 
site would run a job, or to use Condor-G's GlideIn technology.
We have enabled \CondorG to use a matchmaking service to 
automatically select sites for users.

\item We have
extended the ClassAd language, used by the matchmaking framework to describe
resources, to include externally supplied functions to be evaluated at match
time.  This allows the matchmaker to base its decision not only on explicitly
advertised properties but also on opaque logic that is not statically
expressible in a ClassAd. Other uses include incorporation of information
that
is prohibitively expensive to publish in a ClassAd, such as local storage
contents or lists of site-authorized Grid users.  

\item 
We removed the
restriction that the job submission client had to be on the same machine as
the queuing system and enabled the client to securely communicate with the
queue across a network, thus creating a multi-tiered job submission
architecture.

\end{itemize}

Fundamentally, these changes are sufficient to form a multi-user,
multi-site job scheduling system for generic jobs. Thus, a novelty of our
design is that we
use the standard Grid technologies to create a highly reusable framework to the
job scheduling, as opposed to writing our own Resource Broker, which
would be specific to our experiments.

In the remainder of this Section we described higher-level features
for the job management, particularly important for data-intensive
applications.

\subsection{Combination of the Advertised and Queried Information in the MMS
\label{sec:combination}}

\begin{figure*}
\resizebox{\textwidth}{8cm}
{\includegraphics{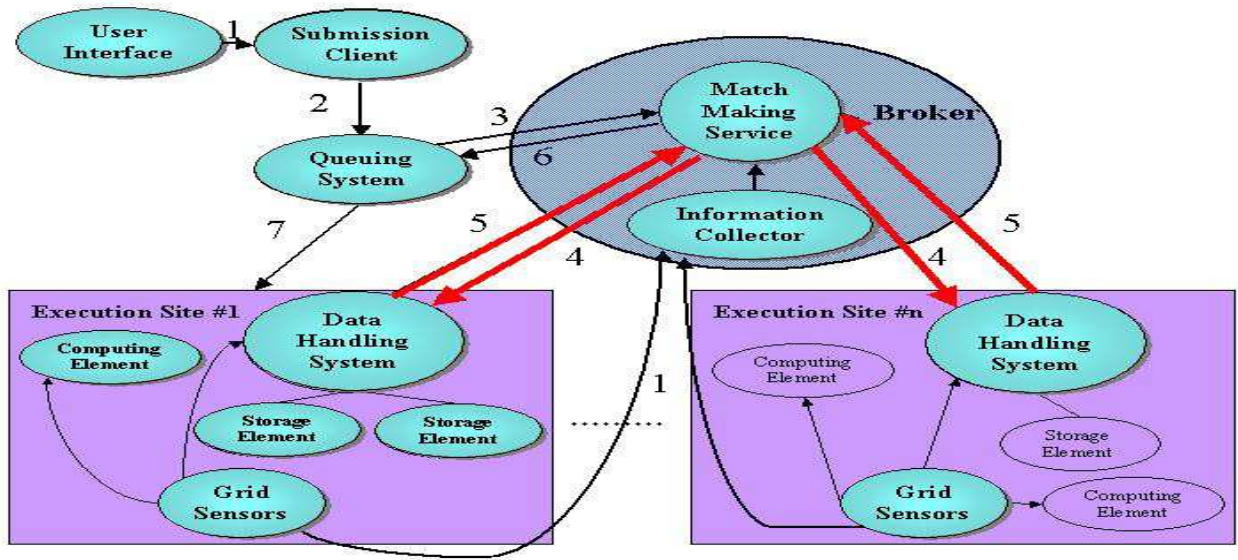}}
\caption{The job management architecture in SAMGrid. 1,2 -- 
Jobs are submitted while resources are advertised, 3 -- MMS matches jobs
with resources, 4,5 -- ranking functions retrieve additional information
from the data handling system, 6,7 -- resource is selected and the job
is scheduled.}
\label{fig:jim_architecture}
\end{figure*}

Classic matchmaking service (MMS) gathers information about the resources
in the form of published ClassAds. This allows
for a general and flexible framework for
resource management (e.g. jobs and resource matching), see \cite{mms}.
There is one important limitation in that scheme, however, which
has to do with the fact that the entities (jobs and resources)
have to be able to express all their relevant properties upfront and 
{\it irrespective of the other party}.

Recall that our primary goal was to enable co-scheduling of jobs
and data. In data-intensive computing, jobs are associated with long
lists of data items (such as files) to be processed by the job.
Similarly, resources are associated with long lists of
data items located, in the network sense, near them. 
For example, jobs requesting thousands of files and sites
having hundreds of thousands of files are not uncommon in production
in the SAM system. Therefore, it would not be scalable to 
explicitly publish all the properties of jobs and resources in the ClassAds.

Furthermore, in order to rank jobs at a resource 
(or resources for the job), we wish to include additional information
that couldn't be expressed in the ClassAds at the time of publishing,
i.e., before the match. Rather, we can analyze such an information
during the match, {\it in the context of the job request}.
For example, a site may prefer a job based on similar
{\it already scheduled} data handling requests. 
\footnote{Unlike the information
about data already placed at sites, 
the information about scheduled data requests, and their estimated
time of completion, is not described by any popular concept like
replica catalog.} Another example of useful additional information, 
not specific to data-intensive computing,
is the pre-authorization of the job's owner with the participating
cite, by means of e.g. looking up the user's grid subject in the 
site's {\bf gridmapfile}. Such a pre-authorization is not a replacement
of security, but rather a means of protecting the matchmaker from
some blunders that otherwise tend to occur in practice.

The original MMS scheme allowed for such
additional information incorporation only in the {\it claiming} phase, i.e.,
after the match when the job's scheduler actually contacts the machine.
In the SAMGrid design, we augment information processing by the MMS 
with the ability
{\it to query} the resources with a job in the context. This is
pictured in Figure \ref{fig:jim_architecture} by arrows extending
from the resource selector to the resources, specifically to the
local data handling agents, in the course of matching. It is implemented
by means of externally supplied ranking function whose evaluation
involves remote call invocation at the resources' sites. Specifically,
the resource ClassAds in our design contain pointers to additional information
providers (data handling servers called Stations): 
\begin{verbatim}
Station_ID = foo
\end{verbatim}
and the body of the MMS ranking function called from the job classAd
\begin{verbatim}
Rank = fun(job_dataset, 
           OTHER.Station_ID)
\end{verbatim}
includes logic similar to this pseudo-code:
\begin{verbatim}
station = resolve(Station_ID,...)
return station->
 get_preference(job_dataset,...)
\end{verbatim}
In the next
subsection, we discuss how we believe this will improve the co-scheduling
of jobs with the data.

\subsection{Interfacing with the SAM Data Handling 
System\label{sec:dh_interface}}

The co-scheduling of jobs and data has always been critical for the SAM system,
where at least a subset of HEP analysis jobs (as of the time of writing,
the dominating class) have their latencies dominated by data access.
Please note that the SAM system already implemented the advanced
feature of retrieving multi-file datasets asynchronously with respect
to the user jobs \cite{hpdc2000,sam_submit} -- this was done initially at the
cluster level rather than at the grid level.

Generally with the data-intensive jobs, we attempt to 
minimize the time to retrieve any missing data and the time
to store output data, as these times propagate into the job's overall
latency. 
As we try to minimize the grid job latency, we
ensure that the design of our system, Figure~\ref{fig:jim_architecture} 
is such that the data handling latencies will be taken into 
account in the process of job matching. This is a principal point
of the present paper, i.e., while we do not yet possess sufficient
real statistics that would justify certain design decisions, we
stress that our system design enables the various strategies and
supports considerations listed below.


In the minimally intelligent implementation, we prefer sites that 
contain most of the the job's data. Our design does not rely on a 
replica catalogue because in the general case, we need 
{\it local} metrics computed by and available from the data handling system:
\begin{itemize}
\item{The network speeds for connections to the sources of any missing data;}
\item{The depths of the queues of data requests for both input and output;}
\item{The network speeds for connections to the nearest destination
of the output files.\footnote{In the SAM system the concept of data routing
is implemented such that the first transfer of an output file is seldom
done directly to the final destination.}}
\end{itemize}
It is important that network speeds be provided by a high-level service
in the data handling rather than by a low-level network sensor, for reasons
similar to those why having e.g. a 56Kbps connection to the ISP
does not necessarily enable one to actually download files 
from the Internet with that speed.

\section {The Management of Configuration and Information in JIM
\label{sec:info}}

Naturally, being able to submit jobs and schedule them
more or less efficiently is necessary but not sufficient
for a Grid system. One has to understand how resources can be described
for such decision making, as well as provide a framework for monitoring
of participating clusters and user jobs.

We consider these and other aspects of {\it information management}
to be closely related to issues of Grid configuration. In the JIM
project, we have developed a 
uniform configuration management framework that allows
for generic grid services instantiation, which in turn gives flexibility
in the design of the Grid as well as inter-operability on the Grid
(see below).

In this Section, we briefly introduce the main configuration
framework and then project it onto the various SAMGrid services 
having to do with information.

\subsection{The Core Configuration Mechanism}

Our main proposal is that grid sites be configured using a site-oriented 
schema, which describes both resources and services,
and that grid instantiation at the sites be derived from these
site configurations. We are not proposing any particular site schema at 
this time, although we hope for the Grid community as a whole 
to arrive at a common schema in the future which will allow reasonable
variations such that various grids are still instantiatable.

\begin{figure*}
\resizebox{\textwidth}{8cm}
{\includegraphics{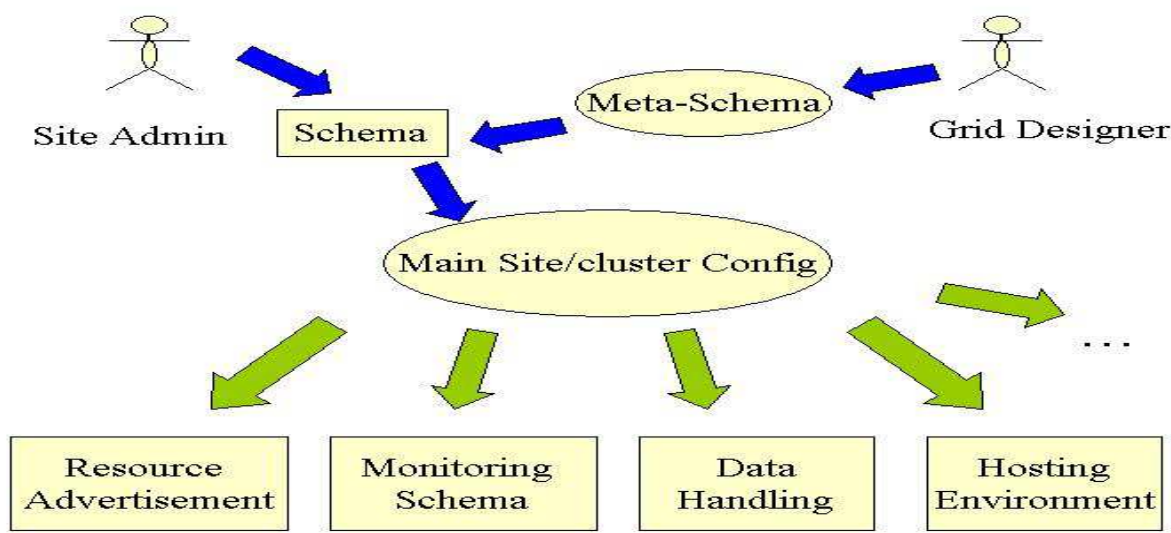}}
\caption{Configuration creation and derivation in our
framework. Service collections are typical of the SAMGrid project.}
\label{fig:fw}
\end{figure*}


Figure \ref{fig:fw} shows configuration derivation in the course
of instantiation of a grid at a site. The site configuration is created
using a meta-configurator similar to one we propose below.

\subsubsection{The Core Meta-Configurator and the Family of Configurators
\label{sec:metaconf}}

In our framework, we create site and {\it all} other configurations 
by a universal tool which we call a {\it meta-configurator}, or
configurator of configurators. The idea is to separate the process of 
querying the user for values of attributes from the {\it schema}
that describes what those attributes are, how they should be queried,
how to guess the default values, and how to derive values of
attributes from those of other attributes. Any concrete configurator
uses a concrete schema to ask the relevant questions to the end user
(site administrator) in order to produce that site's configuration.
Any particular schema is in turn derived from a meta-schema. Thus,
the end configuration can be represented as:
$$C = c(S_d, I_u) = c(c(S_0, I_d), I_u),$$
where $C$ is a particular configuration, $c$ is the configuration operation,
$S_d$ is a particular schema reflecting certain design, 
$S_0$ is the meta-schema, $I_d$ and $I_u$ are the inputs of the designer
and the user, respectively.

In our framework, configurations and schemas are structures of the same
type, which we choose to be trees of nodes each containing a set
of distinct attributes. Our choice has been influenced by the successes
of the XML technologies and, naturally, we use XML for representing
these objects.

To exemplify, assume that in our {\it present} design, a grid {\bf site} 
consists of one or more
{\bf clusters} each having a name and an architecture (homogenous),
as well as exactly one {\bf gatekeeper} for Grid access. Example 
configuration is:

\begin{verbatim}
<?xml version='1.0'?>
<site name='FNAL' 
      schema_version='v0_3'>
  <cluster name='samadams' 
           architecture='Linux'>
    <gatekeeper ...>
  </cluster>
</site>
\end{verbatim}
This configuration was produced by the following schema
\begin{verbatim}
<?xml version='1.0'?>
<site cardinalityMin='1'
      cardinalityMax='1'
      name='inquire-default,FNAL' >
   <cluster  cardinalityMin='1'
     name='set,CLUSTERNAME,inquire'
     architecture=
      'inquire-default,exec,uname'/>
</site>
\end{verbatim}
in an interactive session with the site administrator as follows:
\begin{verbatim}                                                                
What is the name of the site ? [FNAL]: 
 <return>
What is the name of cluster 
  at the site 'FNAL'?  samadams
What is the architecture 
  of cluster 'samadams' [Linux]? 
...
\end{verbatim}

When the schema changes or a new cluster is created at the site, the
administrator merely needs to re-run the tool and answer the simple 
questions again.

\subsection{Resource Advertisement for \CondorG}

In the JIM project, we have designed the grid job management as 
follows. We advertise the 
participating grid clusters to an information collector and grid jobs
are matched \cite{mms} with clusters (resources) based on certain criteria
primarily having to do with the data availability at the sites.
We have implemented this job management using Condor-G \cite{condorg} with
extensions that we have designed together with the Condor team
\cite{acat2002_gg}. 

For the job management to work as described in Section~\ref{sec:jobs},
we need to advertise the clusters together
with the {\bf gatekeepers} as the  means for Condor to 
actually schedule and execute the grid job at the remote site.
Thus, our {\it present} design requires that each 
advertisement contain a cluster, a gatekeeper, 
a SAM station (for jobs actually intending to process data)
and a few other attributes that we omit here. Our advertisement
software then selects from the configuration tree
all patterns containing these attributes and then applies 
a ClassAd generation algorithm to each pattern.

The selection of the subtrees that are ClassAd candidates is
based on the XQuery language. Our queries are generic enough as to 
allow for {\it design evolution}, i.e. to be resilient to some modifications
in the schema. When new attributes are added to an element in the
schema, or when the very structure of the tree changes due to
insertion of a new element, our advertisement service will continue
to advertise these clusters with or without the new information
(depending on how the advertiser itself is configured) but the 
important factor is that this site will continue to be available to
our grid.

For example, assume that one cluster at the site from  
subsection~\ref{sec:metaconf} now has a new grid gatekeeper mechanism
from Globus Toolkit 3, in addition to the old one:
\begin{verbatim}
<?xml version='1.0'?>
<site name='FNAL' 
      schema_version='v0_3'>
  <cluster name='samadams' 
           architecture='Linux'>
    <grid_accesses>
       <gatekeeper ...>
       <gatekeeper-gtk3 ...>
    </grid_accesses>
    ...
\end{verbatim}
Assume further that our particular grid is not yet capable of taking 
advantage of the new middleware and we continue
to be interested in the old {\bf gatekeeper} from each cluster.
Our pattern was such that a {\bf gatekeeper}
is a descendant of the {\bf cluster} so we continue to generate meaningful
ClassAds and match jobs with this site's cluster(s).

\subsection{Monitoring Using Globus MDS}

In addition to advertising (pushing) of resource information
for the purpose of job matching, we deploy Globus MDS-2 for pull-based
retrieval of information about the clusters and activities (jobs
and more, such as data access requests) associated with them.
This allows us to enable web-based monitoring, primarily by
humans, for performance and troubleshooting \cite{acat2002_gg}. 
We introduce (or redefine in the context of our project) concepts
of {\bf cluster}, {\bf station} etc, and map them onto the LDAP
attributes in the OID space assigned to our project (the FNAL
organization, to be exact) by the IANA\cite{iana}. We also create additional
branches for the MDS information tree as to represent our
concepts and their relations. 

We derive the values of the {\it dn}'s on the information tree 
from the site configuration. In this framework,
it is truly straightforward to use
XSLT (or a straight XML-parsing library) to select the names
and other attributes of the relevant pieces of configuration.
For example, if the site has
two clusters defined in the configuration file, our software
will automatically instantiate two branches for the information
tree. Note that the resulting tree may of course be 
distributed within the site as we decide e.g. to run an MDS
server at each cluster, which is a separate degree of freedom.

\subsection{Multiple Grid Instantiation and Inter-Operability
\label{sec:inter-operability}}

We have been mentioning that there are in fact several other grid projects
developing high-level Grid solutions; some of the most noteworthy
include the European Datagrid \cite{edg}, the Crossgrid\cite{xgrid_acat2002},
and The NorduGrid \cite{nordugrid_acat2002}.
Inter-Operability of grids (or of solutions on The Grid if you prefer)
is a well-recognized issue in the community. The
High Energy and Nuclear Physics InterGrid \cite{hisb}
and Grid Inter-Operability \cite{grid_inter_operability_project} projects
are some of the most prominent efforts in this area. 
As we have pointed out in the Introduction, we believe that inter-operability
must include {\it the ability to instantiate and maintain
multiple grid service suites at sites}.


A good example of inter-operability in this sense 
is given by various cooperating Web browsers which all understand
the user's bookmarks, mail preferences etc.. Of course, each browser
may give a different look and feel to its ``bookmarks'' menu, and
otherwise treat them in entirely different ways, yet most browsers
tend to save the bookmarks in the common HTML format, which has
{\it de facto} become the standard for bookmarks.
Our framework, proposed and described in this Section,
is a concrete means to facilitate this aspect of inter-operability.
Multiple grid solutions can be instantiated using a grid-neutral,
site-oriented configuration in an XML-based format. 

We can go one step further and envisage that the various
grids instantiated at a site have
additional, separate configuration spaces that can 
easily be conglomerated into a
{\it grid instantiation database}. In practice, this will allow
the administrators e.g., to list all the Globus gatekeepers
with one simple query.




\section{Integration and Project Status \label{sec:status}}

To provide a complete computing solution for the experiments, 
one must integrate grid-level services with those on the fabric.
Ideally, grid-level scheduling complements, rather than interferes with,
that of local batch systems. Likewise, grid-level monitoring
should provide services that are additional (orthogonal) to those
developed at the fabric's facilities (i.e., monitoring of
clusters' batch systems, storage systems etc.).

Our experiments have customized local environments. CDF has been
successfully using Cluster Analysis Facility (CAF), see
\cite{cdf_acat2002}. D0 has been using MCRunJob \cite{mcrunjob}, a workflow
manager which is also part of the CMS computing insfrastructure.
An important part of the SAMGrid project is to integrate its job and
information services with these environments.

For job management, we have implemented GRAM-compliant job managers
which pass control from Grid-GRAM to each of these two systems
(which in turn are on top of the various batch systems). Likewise,
for the purposes of (job) monitoring, these systems supply information
about their jobs to the XML databases which we deploy on the boundary 
between the Grid and the Fabric. (For resource monitoring, these
advertise their various properties using the frameworks described above).

We delivered a complete, integrated 
prototype of SAMGrid in the Fall of 2002. Our
initial testbed linked 11 sites (5 D0 and 6 CDF) and the basic services
of grid job submission, brokering and monitoring. Our near future
plans include further work on the Grid-Fabric interface and
more features for troubleshooting and error recovery.

\section{Summary \label{sec:summary}}

We have presented the two key components of the SAMGrid, a SAM-based
datagrid being used by the Run II experiments at FNAL. To the data
handling capabilities of SAM, we add grid job scheduling and brokering, as
well as information processing and monitoring. We use the standard \CondorG
middleware so as to maximize the reusability of our design. As to the
information management, we have developed a unified framework 
for configuration management in XML, from where we explore
resource advertisement, monitoring and other directions such as
service instantiation. We are deploying SAMGrid at the time of writing
this paper and learning from the new experiences.

\section{Acknowledgements}

This work is sponsored in part by DOE contract No. DE-AC02-76CH03000.
Our collaboration takes place as part of the DOC Particle Physics Data
Grid (PPDG), \cite{ppdg} Collaboratory SciDAC project. We thank
the many participants from the Fermilab Computing Division,
the D0 and CDF Experiments, and the members of the Condor team
for fruitful discussions as well as the development of our 
software and other software that we rely upon.

\end{document}